\documentclass[10pt,a4paper,twocolumn]{article}
\usepackage[lined,algonl,ruled]{algorithm2e}
\usepackage[latin1]{inputenc}
\usepackage{epsfig}
\usepackage{subfigure}
\usepackage{latexsym}
\usepackage{amssymb}
\usepackage{pifont}
\usepackage{SSI}
\usepackage[open-square,italic-theorems]{QED}
\usepackage[a4paper, margin=1.5cm]{geometry}

\newtheorem{thm}{Theorem}[section]
\newtheorem{lem}[thm]{Lemma}

\newtheorem{defn}[thm]{Definition}

\begin{document}

\title{\normalsize{{\bf MISCONFIGURATION MANAGEMENT\\
      OF NETWORK  SECURITY COMPONENTS}}}

\author{
  Fr\'ed\'eric Cuppens$^1$ \hspace{1cm} Nora Cuppens-Boulahia$^1$\\
  \\
  $^1$~GET/ENST-Bretagne,\\
  2, rue de la Ch\^ataigneraie,\\
  35576 Cesson  S\'evign\'e - France\\
  \{frederic.cuppens,nora.cuppens\}@enst-bretagne.fr
  \and
  Joaqu\'in Garc\'ia-Alfaro$^{1,2}$\\
  \\
  $^2$~dEIC/UAB,\\
  Edifici Q, Campus de Bellaterra,\\
  08193, Bellaterra, Barcelona - Spain\\
  joaquin.garcia@uab.es
}

\date{}

\abstract{
  \noindent
  Many companies and organizations use firewalls to control the access
  to their network infrastructure. Firewalls are network security
  components which provide means to filter traffic within corporate
  networks, as well as to police incoming and outcoming interaction
  with the Internet. For this purpose, it is necessary to configure
  firewalls with a set of filtering rules. Nevertheless, the existence
  of errors in a set of filtering rules is very likely to degrade the
  network security policy. The discovering and removal of these
  configuration errors is a serious and complex problem to solve. In
  this paper, we present a set of algorithms for such a management.
  Our approach is based on the analysis of relationships between the
  set of filtering rules. Then, a subsequent rewriting of rules will
  derive from an initial firewall setup -- potentially misconfigured
  -- to an equivalent one completely free of errors. At the same time,
  the algorithms will detect useless rules in the initial firewall
  configuration.
}

\keys{Network Security, Firewalls, Filtering Rules, Redundancy and
  Shadowing of Rules}

\maketitle

\section{Introduction}
\label{sec:introduction}

\begin{table*}[tbp]
\begin{center}
\begin{tabular}{|c|ccccc|c|}
\hline
\textbf{order}&&&\textbf{condition}&&& \textbf{decision} \cr
\hline
&(p)rotocol & (s)ource & (sP)ort & (d)estination & (dP)ort  &  \cr
\hline
1&any & xxx.xxx.xxx.[001,030] & any & xxx.xxx.xxx.[020,045] & any  & deny \cr
2&any & xxx.xxx.xxx.[020,060] & any & xxx.xxx.xxx.[025,035] & any  & accept \cr
3&any & xxx.xxx.xxx.[040,070] & any & xxx.xxx.xxx.[020,045] & any  & accept \cr
4&any & xxx.xxx.xxx.[015,045] & any & xxx.xxx.xxx.[025,030] & any  & deny \cr
5&any & xxx.xxx.xxx.[025,045] & any & xxx.xxx.xxx.[020,040] & any  & accept \cr
\hline
\end{tabular}
\caption{Example of a set of filtering rules with five condition attributes.}
\label{tab:frules}
\end{center}
\end{table*}

The use of firewalls is the dominant method for companies and
organizations to segment access control within their own networks.
They are typically deployed to filter traffic between \textit{trusted}
and \textit{untrusted} zones of corporate networks, as well as to
police their incoming and outcoming interaction with the
Internet\footnote{Firewalls also implement other functionalities, such
  as Proxying and Network Address Transfer (NAT), but it is not the
  purpose of this paper to cover these functionalities.}.

Firewalls are network security components, with several interfaces
associated with the different zones of the network. A company may
partition, for instance, its network into three different zones: a
demilitarized zone (DMZ for short), a private network and a zone for
security administration. In this case, one may use a firewall with
three interfaces associated with these three zones, as well as a
fourth interface to control the access to the Internet.

In order to apply the filtering process, it is necessary to configure
the firewall with a set of filtering rules (e.g.,~the set of filtering
rules shown in Table \ref{tab:frules}). Each filtering rule typically
specifies a $decision$ (e.g.,~$accept$ or $deny$) that applies to a
set of $condition$ attributes, such as protocol, source, destination,
and so on.

For our work, we define a filtering rule as follows:

\vspace*{-.25cm}
\begin{equation}
  R_i: \{condition_i\} \rightarrow decision_i
\end{equation}

where $i$ is the relative position of the rule within the set of
rules, $decision_i$ is a boolean expression in
$\{accept,deny\}$\footnote{The $decision$ field may also be a
  combination of both $accept$ and $deny$ together with some other
  options such as a logging or jump options. For reasons of clarity we
  assume that just accept and deny are proper values.}, and
$\{condition_i\}$ is a conjunctive set of condition attributes such
that $\{condition_i\}$ equals $A_1 \wedge A_2 \wedge ... \wedge A_p$,
and $p$ is the number of condition attributes of the given filtering
rules.

The following example\footnote{To simplify the example, the number of
  condition attributes, i.e., $p$, is just two: (s)ource and
  (d)destination. We do not show the condition attributes (p)rotocol,
  (sP)ort, and (dP)ort, because their value will always be $true$.}
shows the filtering rules of Table \ref{tab:frules} using such a
formalism.

\begin{center}
\begin{minipage}{6.5cm}
$R_1: (s \in [1,30] \land d \in [20,45]) \rightarrow deny$ \\
$R_2: (s \in [20,60] \land d \in [25,35]) \rightarrow accept$ \\
$R_3: (s \in [40,70] \land d \in [20,45]) \rightarrow accept$ \\
$R_4: (s \in [15,45] \land d \in [25,30]) \rightarrow deny$ \\
$R_5: (s \in [25,45] \land d \in [20,40]) \rightarrow accept$
\end{minipage}
\end{center}

\begin{figure}[htbp]
  \centering
  \epsfig{file=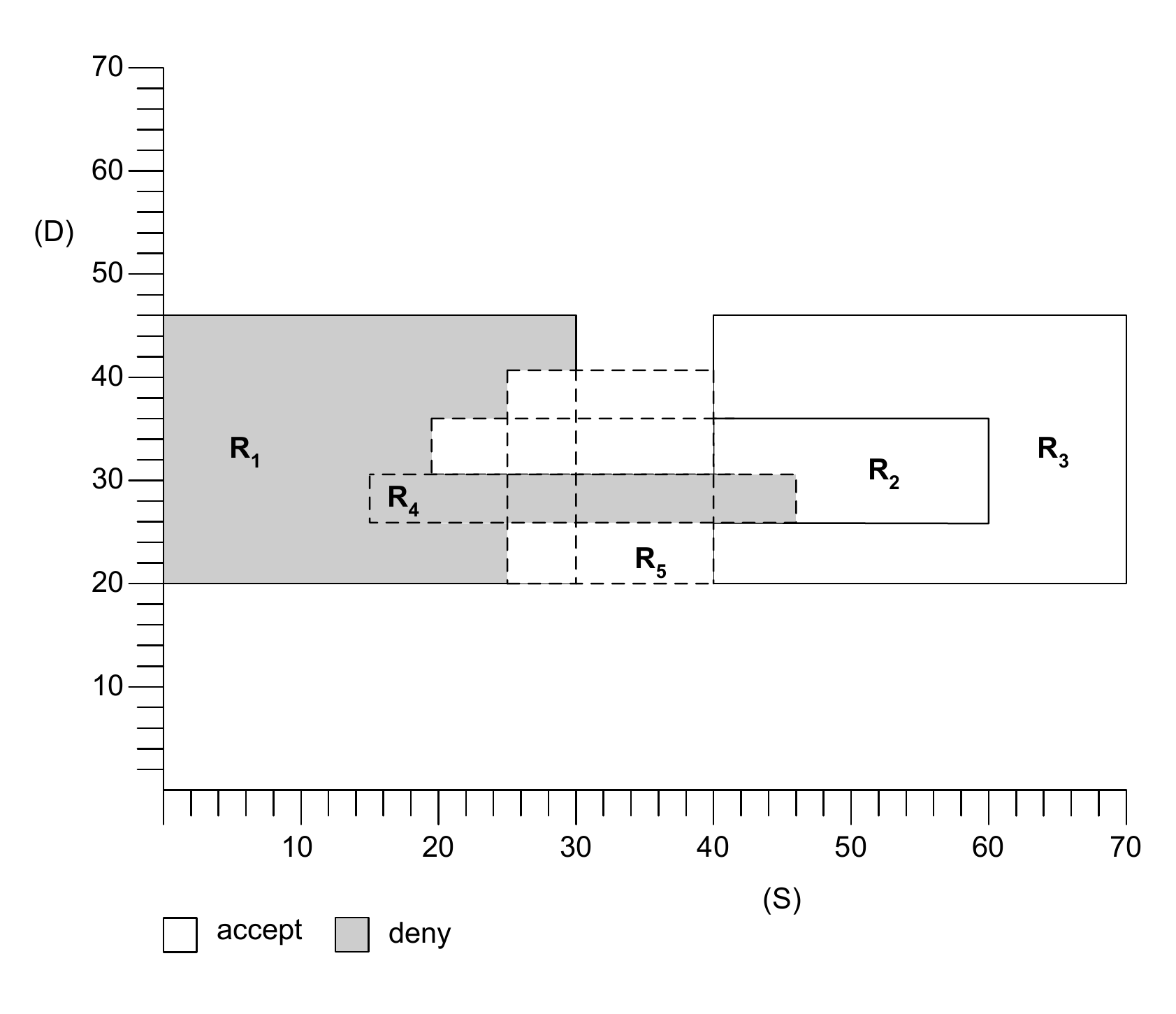,width=6.5cm}
  \vspace{-.15cm}
  \caption{Main overlaps within the rules of Table \ref{tab:frules}}
  \label{fig:mitivationExample}
\end{figure}

When processing packages, conflicts due to rule overlaps can occur
within the filtering policy. For instance, we can see in Figure
\ref{fig:mitivationExample} a geometrical representation of the main
overlaps within the filtering rules of Table \ref{tab:frules}.

To solve these conflicts, most firewall implementations use a
\textit{first matching} strategy through the ordering of rules -- such
as the \textit{order} field shown in Table \ref{tab:frules}. This way,
each packet processed by the firewall is mapped to the decision of the
rule with highest priority. This strategy introduces, however, new
configuration errors, such as \textit{shadowing} of rules and
\textit{redundancy}. For our work, we define these two general cases
of firewall misconfiguration as follows.

\begin{defn}
\label{def:shadowing}
Let $R$ be a set of filtering rules. Then $R$ has \textbf{shadowing}
if and only if there exists at least one filtering rule, $R_i$ in $R$,
which never applies because all the packets that $R_i$ may match, are
previously matched by another rule, or combination of rules, with
higher priority in order.
\end{defn}

\begin{defn}
\label{def:redundancy}
Let $R$ be a set of filtering rules. Then $R$ has \textbf{redundancy}
if and only if there exists at least one filtering rule, $R_i$ in $R$,
such that the following conditions hold: (1) $R_i$ is not shadowed by
any other rule; (2) when removing $R_i$ from $R$, the filtering
result, i.e., the security policy, does not change.
\end{defn}

The discovering and removal of both redundancy and shadowing is a
serious problem which must be solved since a misconfigured set of
filtering rules, if not handled correctly, is very likely to cause
packets to be subject to the wrong actions, and to lead to a weak
security policy.

In this paper, we present a set of algorithms for the discovering and
removal of both redundancy and shadowing of rules. Our main objective
is the following. Given a specific firewall setup, we want to analyze
the existing firewall configuration to check whether there is errors
in such a configuration, i.e., the set of filtering rules presents
shadowing or redundancy as defined above.

Our approach is based on the relationships between the filtering
rules' parameters: coincidence, disjunction and inclusion. We use a
rule transformation process that derive from a set of filtering rules
to an equivalent and valid one that is completely free of both
shadowing and redundancy.

The advantages of our proposal are threefold. First of all, after
rewriting the rules one can verify that there is no redundancy nor
shadowing in the resulting firewall configuration. Each redundant or
shadowed rule -- considered as useless during the audit process --
will be removed from the initial set of filtering rules.

On the second hand, when such a detection occurs the discovering
process will provide an evidence of error to the administration
console. This way, the security officer in charge of the network can
check from the initial specification, in order to verify the
correctness of the whole process.

On the third hand, the resulting rules are completely disjoint, i.e.,
the ordering of rules is no longer relevant. Hence, one can perform a
second transformation in a positive or negative manner: positive, when
generating only permissions; and negative, when generating only
prohibitions. Positive rewriting can be used in a closed policy
whereas negative rewriting can be used in case of an open policy.

After performing this second rewriting, the security officer will
have a clear view of the accepted traffic (in the case of positive
rewriting) or the rejected traffic (in the case of negative
rewriting).

The rest of this paper is organized as follows: Section
\ref{sec:relatedWork} starts with an analysis of some related work.
Then, Section \ref{sec:proposal} presents our algorithms and
introduces some examples to validate the correctness of our approach.
Section \ref{sec:evaluation} analyzes the complexity of our proposed
algorithms and overviews a performance study.
Section~\ref{sec:conclusions} closes the paper with some conclusions
and gives an outlook on future work.

\section{Related Work}
\label{sec:relatedWork}

A first approach to get a firewall configuration free of errors is by
applying a formal security model to express the network security
policy. In \cite{fast}, for example, a formal model is presented with
this purpose. This way, a set of filtering rules, whose syntax is
specific to a given firewall, may be generated using a transformation
language. Nonetheless, this approach is not enough to ensure that the
firewall configuration is completely free of errors.

Some other proposals, such as \cite{adiseshu00, gupta00, al-shaer04,
  liu04-fwqueries, al-shaer05}, provide means to directly manage
misconfiguration. For instance, the authors in \cite{adiseshu00}
consider that, in a configuration set, two rules are in conflict when
the first rule in order matches some packets that match the second
rule, and the second rule also matches some of the packets that match
the first rule.

This approach is very limited since it does not detect what we
consider \textit{serious misconfiguration errors}, i.e., redundancy
and shadowing of rules (cf.~Section\ref{sec:introduction},
Def.~\ref{def:shadowing} and Def.~\ref{def:redundancy}). What they
detect is just a particular case of wrongly defined rules which cause
ambiguity in the firewall configuration, and that is more efficiently
defined as a combination of both redundancy and shadowing.

In \cite{gupta00}, two new cases of misconfiguration are considered.
First, a rule $R_j$ is defined as backward redundant if and only if
there exists another rule $R_i$ with higher priority in order such
that all the packets that match rule $R_j$ also match rule $R_i$. On
the other hand, a rule $R_i$ is defined as forward redundant if and
only if there exists another rule $R_j$ with the same decision and
less priority in order such that the following conditions hold: (1)
all the packets that match $R_i$ also match $R_j$; (2) for each rule
$R_k$ between $R_i$ and $R_j$, and that matches all the packets that
also match rule $R_i$, $R_k$ has the same decision as $R_i$.

Although this approach seems to head in the right direction, we
consider our definitions (cf.~Section\ref{sec:introduction},
Def.~\ref{def:shadowing} and Def.~\ref{def:redundancy}) simpler and
more general, because all possible backward and forward redundant
rules are specific cases of both redundancy and shadowing, but not
vice versa. For instance, given the following set of rules:
\begin{center}
  \vspace{-0.15cm}
\begin{minipage}{4.5cm}
$R_1: s \in [10,50] \rightarrow deny$ \\
$R_2: s \in [40,70] \rightarrow accept$\\
$R_3: s \in [50,80] \rightarrow accept$
\end{minipage}
\vspace{-0.15cm}
\end{center}
Since rule $R_2$ comes after rule $R_1$, rule $R_2$ only applies over
the interval $[51,70]$ -- i.e., $R_2$ is redundant with respect to
rule $R_3$. Their detection proposal, as defined above, cannot detect
the redundancy of rule $R_2$. Therefore, we point out this work as
incomplete.

To our best knowledge, the authors of the \textit{firewall policy
  advisor} \cite{al-shaer04, al-shaer05} propose the most efficient
set of techniques and algorithms to detect redundancy and shadowing in
different firewall configuration setups. In addition to the discovery
process, their approach also attempts an optimal insertion of
arbitrary rules into an existing configuration, through a tree based
representation of the filtering criteria.

Nonetheless, and even though the efficiency of their proposed
discovering algorithms and techniques is very promising, we also
consider this approach as incomplete.

On the one hand, their approach is too weak since, given a
misconfigured firewall, their discovering algorithms could not detect
all the possible errors. For example, given the following set of
rules:
\begin{center}
\vspace{-0.15cm}
\begin{minipage}{4.5cm}
$R_1: s \in [10,50] \rightarrow accept$ \\
$R_2: s \in [40,90] \rightarrow accept$ \\
$R_3: s \in [30,80] \rightarrow deny$
\end{minipage}
\vspace{-0.15cm}
\end{center}
their approach cannot detect the shadowing over rule $R_3$ due to the
union of rules $R_1$ and $R_2$.

On the other hand, the authors do not cover, intentionally, an
automatic rewriting of rules to correct the discovered errors. This
way, it is the security officer who should perform the final changes.

Summing up, we believe that none of the identified related work
provides a complete discovering of both redundancy and shadowing of
rules -- which are the cases we consider \textit{serious errors}
within firewalls configurations -- as well as a proper handling of
such a misconfiguration.

\section{Proposed Algorithms}
\label{sec:proposal}

\subsection{Detection Process}
\label{sec:detectingShadowing}

As pointed out in Section \ref{sec:introduction}, our main objective
is the discovering of both shadowing and redundancy errors inside an
initial set of filtering rules $R$. Such a detection process is a way
to alert the security officer in charge of the network about these
configuration errors, as well as to remove all the useless rules in
the initial firewall configuration.

The data to be used for the detection process is the following. A set
of rules $R$ as a dynamic linked-list\footnote{A dynamic linked-list
  is a pointer-based data structure that can be used to properly
  represent the abstract notion of a dynamic list.} of initial size
$n$, where $n$ equals $count(R)$, and where each element is an
associative array\footnote{Associative arrays -- also known as a map,
  lookup table, or dictionary -- have strings as keys and behave more
  like two-column tables, where the first column is the key to access
  the value of the second column.} with the strings $condition$,
$decision$, $shadowing$, and $redundancy$ as keys to access each
necessary value.

To simplify, we assume one can access a linked-list through the
operator $R_i$, where $i$ is the relative position regarding the
initial list size -- $count(R)$. We also assume one can add new values
to the list as any other normal variable does ($element \leftarrow
value$), as well as to remove elements through the addition of an
empty set ($element \leftarrow \emptyset$). The internal order of
elements from the linked-list $R$ keeps with the relative ordering of
rules.

In turn, each element $R_i[condition]$ is an indexed
array\footnote{For our algorithms, we assume that the keys of an
  indexed array are integers, beginning at 1, and where one can
  identify the elements by their position.} of size $p$ containing the
set of conditions of each rule; each element $R_i[decision]$ is a
boolean variable whose values are in $\{accept,deny\}$; each element
$R_i[shadowing]$ is a boolean variable in $\{true,false\}$; each
element $R_i[redundancy]$ is another boolean variable in
$\{true,false\}$. Both shadowing and redundancy variables of each rule
are initialized to $false$ by default.

For reasons of clarity, we split the whole detection process and the
removal of misconfiguration in two different processes. Thus, we
define a main detection function (Algorithm \ref{alg:detection}),
whose input is the initial set of filtering rules, $R$, and an
auxiliary function (Algorithm \ref{alg:exclusion}) whose input is two
rules, $A$ and $B$. Once executed, this auxiliary function returns a
further rule, $C$, whose set of condition attributes is the exclusion
of the set of conditions from $A$ over $B$. In order to simplify the
representation of this second algorithm (cf.~Algorithm
\ref{alg:exclusion}), we use the notation $A_i$ as an abbreviation of
the variable $A[condition][i]$, an the notation $B_i$ as an
abbreviation of the variable $B[condition][i]$ -- where $i$ in
$[1,p]$.

We recall that the output of the main detection function is the set
which results as a transformation of the initial set $R$. This new set
is equivalent to the initial one, $R$, and all its rules are
completely disjoint. Therefore, the resulting set is free of both
redundancy and shadowing of rules, as well as any other possible
configuration error.

\begin{algorithm}
  \caption{\texttt{detection}($R$)}
  \label{alg:detection}
  \SetKwFunction{Exclusion}{exclusion}
  \Begin{
    \For{$i\leftarrow 1$ \KwTo $(count(R)-1)$\\}{
      \For{$j\leftarrow (i+1)$ \KwTo $count(R)$\\}{
        $R_j \leftarrow$ \Exclusion($R_j$,$R_i$);\\
        \If{$R_j[condition] = \emptyset$\\}{
          $R_j[shadowing] \leftarrow true$;
        }
      }
    }
  }
\end{algorithm}

\begin{algorithm}
\caption{\texttt{exclusion}($B$,$A$)}
\label{alg:exclusion}
  \Begin{
    $C[condition] \leftarrow \emptyset$;\\
    $C[decision] \leftarrow B[decision]$;\\
    $C[shadowing] \leftarrow false$;\\
    $C[redundancy] \leftarrow false$;\\
    \ForAll{\small{\textbf{the elements of} $A[condition]$
        \textbf{and} $B[condition]$}}{
        \eIf{$((A_1 \cap B_1) \neq \emptyset$ {\bf and} \\
          $(A_2 \cap B_2) \neq \emptyset$ {\bf and} \ldots\\
          ~~~~~\ldots~~~~{\bf and} $(A_p \cap B_p) \neq
          \emptyset)$\\}{
          $C[condition] \leftarrow C[condition]~\cup$ \\
          \footnotesize{
            \{$(B_1 - A_1) \wedge B_2 \wedge ... \wedge B_p$,\\
            $(A_1 \cap B_1) \wedge (B_2 - A_2) \wedge ... \wedge B_p$,\\
            $(A_1 \cap B_1) \wedge (A_2 \cap B_2) \wedge (B_3 - A_3) \wedge ... \wedge B_p$,\\
            $ ... $\\
            $(A_1 \cap B_1) \wedge ... \wedge (A_{p-1} \cap B_{p-1})
            \wedge (B_p -
            A_p) \}$;\\
          } }{
          $C[condition] \leftarrow$ \\
          ~~$(C[condition] \cup B[condition]$); } } \Return{ $C$; } }
\end{algorithm}

\subsubsection{Applying the Algorithms}

This section gives a short outlook on applying algorithms
\ref{alg:detection} and \ref{alg:exclusion} over some
representative examples.

Let us start applying the function \textit{exclusion} (Algorithm
\ref{alg:exclusion}) over a set of two rules $R_i$ and $R_j$, each one
of them with two condition attributes -- (s)ource and (d)estination --
and where rule $R_j$ has less priority in order than rule $R_i$. In
this first example:
\begin{center}
\begin{minipage}{7.25cm}
  $R_i[condition] = (s \in [80,100]) \wedge (d \in [1,50])$\\
  $R_j[condition] = (s \in [1,50]) \wedge (d \in [1,50])$
\end{minipage}
\end{center}

since $(s \in [1,50]) \cap (s \in [80,100])$ equals $\emptyset$, the
condition attributes of rules $R_i$ and $R_j$ are completely
independent. Thus, the applying of $exclusion(R_j,R_i)$ is equal to
$R_j[condition]$.

The following three examples show the same execution over a set of
condition attributes with different cases of conflict. A first case is
the following:
\begin{center}
\begin{minipage}{7.25cm}
  $R_i[condition] = (s \in [1,60]) \wedge (d \in [1,30])$\\
  $R_j[condition] = (s \in [1,50]) \wedge (d \in [1,50])$
\end{minipage}
\end{center}
where there is a main overlap of attribute $s$ from $R_i[condition]$
which completely excludes the same attribute on $R_j[condition]$.
Then, there is a second overlap of attribute $d$ from $R_i[condition]$
which partially excludes the range $[1,30]$ into attribute $d$ of
$R_j[condition]$, which becomes $d$ in $[31,50]$. This way,
$exclusion(R_j,R_i) \leftarrow \{(s \in [1,50]) \wedge (d \in [31,50])
\}$\footnote{For reasons of clarity, we do not show the first empty
  set corresponding to the first overlap. If shown, the result should
  become as follows: $exclusion(R_j,R_i) \leftarrow \{\emptyset, (s
  \in [1,50]) \wedge (d \in [31,50]) \}$.}. In this other example:
\begin{center}
\begin{minipage}{7.25cm}
  $R_i[condition] = (s \in [1,60]) \wedge (d \in [20,30])$\\
  $R_j[condition] = (s \in [1,50]) \wedge (d \in [1,50])$
\end{minipage}
\end{center}
there is two simple overlaps of both attributes $s$ and $d$ from
$R_i[condition]$ to $R_j[condition]$, such that $exclusion(R_j,R_i)$
becomes $\{(s \in [1,50]) \wedge (d \in [1,19]), (s \in [1,50]) \wedge
(d \in [31,50]) \}$.\\

A more complete example is the following,
\begin{center}
\begin{minipage}{7.25cm}
  $R_i[condition] = (s \in [10,40]) \wedge (d \in [20,30])$\\
  $R_j[condition] = (s \in [1,50]) \wedge (d \in [1,50])$
\end{minipage}
\end{center}
where $exclusion(R_j,R_i)$ becomes $\{ (s \in [1,9]) \wedge (d \in
[1,50]), (s \in [41,50]) \wedge (d \in [1,50]), (s \in [10,40]) \wedge
(d \in [1,19]), (s \in [10,40]) \wedge (d \in [31,50]) \}$.

Regarding a full exclusion, let us show the following example,
\begin{center}
\begin{minipage}{7.25cm}
  $R_i[condition] = (s \in [1,60]) \wedge (d \in [1,60])$\\
  $R_j[condition] = (s \in [1,50]) \wedge (d \in [1,50])$
\end{minipage}
\end{center}
where the set of condition attributes of rule $R_i$ completely
excludes the ones of rule $R_j$. Then, the applying of
$exclusion(R_j,R_i)$ becomes an empty set (i.e.,
$\{\emptyset,\emptyset\} = \emptyset$). Hence, on a further execution
of Algorithm \ref{alg:detection} the shadowing field of rule $R_j$
(initialized as $false$ by default) would become $true$ (i.e.,
$R_j[shadowing] \leftarrow true$).

To conclude this section, let us show a complete execution of
algorithms \ref{alg:detection} and \ref{alg:exclusion} over a set of
filtering rules based on Table \ref{tab:frules} -- whose main overlaps
have been previously shown in Figure \ref{fig:mitivationExample}.

{\small
\begin{center}
\framebox[7cm][l]{
\begin{minipage}{7cm}
  $/*motivation~example*/$\\
  ~~\\
  $R_1: (s \in [1,30] \land d \in [20,45]) \rightarrow deny$ \\
  $R_2: (s \in [20,60] \land d \in [25,35]) \rightarrow accept$ \\
  $R_3: (s \in [40,70] \land d \in [20,45]) \rightarrow accept$ \\
  $R_4: (s \in [15,45] \land d \in [25,30]) \rightarrow deny$ \\
  $R_5: (s \in [25,45] \land d \in [20,40]) \rightarrow accept$
\end{minipage} }
\framebox[7cm][l]{
\begin{minipage}{7cm}
  $/*step~1*/$\\
  ~~\\
  $R_1: (s \in [1,30] \land d \in [20,45]) \rightarrow deny$ \\
  $R_2: (s \in [31,60] \land d \in [25,35]) \rightarrow accept$ \\
  $R_3: (s \in [40,70] \land d \in [20,45]) \rightarrow accept$ \\
  $R_4: (s \in [31,45] \land d \in [25,30]) \rightarrow deny$ \\
  $R_5: (s \in [31,45] \land d \in [20,40]) \rightarrow accept$
\end{minipage} }
\framebox[7cm][l]{
\begin{minipage}{7cm}
  $/*step~2*/$\\
  ~~\\
  $R_1: (s \in [1,30] \land d \in [20,45]) \rightarrow deny$ \\
  $R_2: (s \in [31,60] \land d \in [25,35]) \rightarrow accept$ \\
  $R_3: \{(s \in [61,70] \land d \in [20,45]),\\
  ~~~~~~~~~(s \in [40,60] \land d \in [20,24]),\\
  ~~~~~~~~~(s \in [40,60] \land d \in [36,45])\} \rightarrow accept$ \\
  $R_4: \emptyset \rightarrow deny$\\
  $R_5: \{(s \in [31,45] \land d \in [20,24]),\\
  ~~~~~~~~~(s \in [31,45] \land d \in [36,40])\} \rightarrow accept$
\end{minipage} }
\framebox[7cm][l]{
\begin{minipage}{7cm}
  $/*step~3=step~4=resulting~rules*/$\\
  ~~\\
  $R_1: (s \in [1,30] \land d \in [20,45]) \rightarrow deny$ \\
  $R_2: (s \in [31,60] \land d \in [25,35]) \rightarrow accept$ \\
  $R_3: \{(s \in [61,70] \land d \in [20,45]),\\
  ~~~~~~~~~(s \in [40,60] \land d \in [20,24]),\\
  ~~~~~~~~~(s \in [40,60] \land d \in [36,45])\} \rightarrow accept$ \\
  $R_5: \{(s \in [31,39] \land d \in [20,24]),\\
  ~~~~~~~~~(s \in [31,39] \land d \in [36,40])\} \rightarrow accept$
\end{minipage} }
\framebox[7cm][l]{
\begin{minipage}{7cm}
  $/*warnings*/$\\
  ~~\\
  $R_4[shadowing] = true$
\end{minipage} }
\end{center}
}

\subsection{Correctness of the Algorithms}

\begin{defn}\label{def:set}
  Let $R$ be a set of filtering rules and let $Tr(R)$ be the resulting
  filtering rules obtained by applying Algorithm
  \ref{alg:detection} to $R$.
\end{defn}

\begin{lem}\label{lem:equivalence}
  Let $R_i: condition_i \rightarrow decision_i$ and $R_j: condition_j
  \rightarrow decision_j$ be two filtering rules. Then $\{R_i,R_j\}$
  is equivalent to $\{R_i,R'_j\}$ where $R'_j \leftarrow
  exclusion(R_j, R_i)$.\footnote{A set of proofs to validate the
    theorems and lemmas of this section is provided in
    \cite{phoenix}.}
\end{lem}

\begin{thm}\label{thm:equivalence}
  Let $R$ be a set of filtering rules and let $Tr(R)$ be the resulting
  filtering rules obtained by applying Algorithm
  \ref{alg:detection} to $R$. Then $R$ and $Tr(R)$ are
  equivalent.
\end{thm}

\begin{lem}\label{lem:simultaneousness}
  Let $R_i: condition_i \rightarrow decision_i$ and $R_j: condition_j
  \rightarrow decision_j$ be two filtering rules. Then rules $R_i$ and
  $R'_j$, where $R'_j \leftarrow exclusion(R_j, R_i)$ will never
  simultaneously apply to any given packet.
\end{lem}

\begin{thm}\label{thm:ordering}
  Let $R$ be a set of filtering rules and let $Tr(R)$ be the resulting
  filtering rules obtained by applying Algorithm \ref{alg:detection}
  to $R$. Then ordering the rules in $Tr(R)$ is no longer relevant.
\end{thm}

\begin{thm}\label{thm:freedom}
  Let $R$ be a set of filtering rules and let $Tr(R)$ be the resulting
  filtering rules obtained by applying Algorithm
  \ref{alg:detection} to $R$. Then $Tr(R)$ is free from both
  shadowing and redundancy.
\end{thm}

\subsection{Complete Detection}
\label{sec:extendedDetection}

Up to now, the result of Algorithm \ref{alg:detection} offers a set of
filtering rules, $Tr(R)$, equivalent to an initial set of rules $R$,
and completely free of any possible relation between its rules.
Nevertheless, there is a limitation on such an algorithm regarding the
reporting of redundancy -- just the existence of shadowing is reported
to the security officer. Therefore, we need to modify this algorithm
in order to also detect redundancy in $R$.

The purpose of this section is to solve this limitation, by presenting
a second manner to completely discover both shadowing and redundancy
errors into the initial set of filtering rules, $R$, based on the
techniques and results previously shown in Section
\ref{sec:detectingShadowing}.

\begin{algorithm}
  \caption{\texttt{testRedundancy}($R$,$i$)}
  \label{alg:testRedundancy}
  \SetKwFunction{Exclusion}{exclusion}
  \Begin{
    $test \leftarrow false$;\\
    $j \leftarrow (i+1)$;\\
    $temp \leftarrow R_i$;\\
    \While{$\neg test$ {\bf and} ($j \leq count(R)$)\\}{
      \If{$temp[decision] = R_j[decision]$\\}{
        $temp \leftarrow \Exclusion(temp,R_j)$;\\
        \If{temp[condition] = $\emptyset$\\}{
          $test \leftarrow true$;\\
        } }
      $j \leftarrow (j+1)$;\\
    } \Return{ $test$; } }
\end{algorithm}

\begin{algorithm}
  \caption{\texttt{completeDetection}($R$)}
  \label{alg:complete}
  \SetKwFunction{Exclusion}{exclusion}
  \SetKwFunction{TestRedundancy}{testRedundancy}
  \Begin{
    \tcc{ Phase 1 }
    \For{$i \leftarrow 1$ \KwTo $(count(R)-1)$}{
        \For{$j \leftarrow (i+1)$ \KwTo $count(R)$}{
          \lIf{$R_i[decision] \neq R_j[decision]$}{\\
            ~~$R_j \leftarrow$ \Exclusion($R_j$,$R_i$);\\
          }
          \lIf{$R_j[condition] = \emptyset$}{\\
            ~~$R_j[shadowing] \leftarrow true$;
          }
        }
    }
    \tcc{ Phase 2 }
    \For{$i \leftarrow 1$ \KwTo $(count(R)-1)$}{
        \eIf{\TestRedundancy($R,i$)}{
          $R_i[condition] \leftarrow \emptyset$;
          $R_i[redundancy] \leftarrow true$;
        }{
          \For{$j \leftarrow (i+1)$ \KwTo $count(R)$}{
            \lIf{$R_i$[decision]=$R_j$[decision]}{\\
              ~~$R_j\leftarrow$\Exclusion($R_j$,$R_i$);\\
            }
            \lIf{($\neg R_j[redundancy]$ \textbf{and}\\
              $R_j$[condition] = $\emptyset$)}{\\
              ~~$R_j[shadowing] \leftarrow true$; } } }
    }
  }
\end{algorithm}

The reporting of redundancy is much more complex than the task of
reporting shadowing. To properly overcome this complexity, we first
divide the whole process in two different algorithms (Algorithm
\ref{alg:testRedundancy} and Algorithm \ref{alg:complete}).

The first algorithm (cf.~Algorithm \ref{alg:testRedundancy}) is a
boolean function in $\{true,false\}$, which, in turn, applies the
transformation \textit{exclusion} (cf.~Section
\ref{sec:detectingShadowing}, Algorithm \ref{alg:exclusion}) over a
set of filtering rules to check whether the rule obtained as a
parameter is potentially redundant.

The second algorithm (cf.~Algorithm \ref{alg:complete}) performs the
whole process of detecting and removing both redundancy and shadowing,
and is also split in two different phases. During the first phase, a
set of shadowing rules are detected and removed from a top-bottom
scope, by iteratively applying Algorithm \ref{alg:exclusion} -- when
the decision field of the two rules is different. Let us notice that
this stage of detecting and removing shadowed rules is applied before
the detection and removal of proper redundant rules.

The resulting set of rules is then used when applying the second
phase, also from a top-bottom scope. This stage is performed to detect
and remove proper redundant rules, as well as to detect and remove all
the further shadowed rules resulting during the latter process.

As a result of the whole execution, the initial set of rules, $R$, is
transformed into an equivalent set, $Tr(R)$, whose rules are
completely disjoint. Furthermore, all the discovery of both shadowing
and redundancy is reported to the security officer, who may verify the
whole process.

\subsubsection{Applying the Algorithms}

In this section we give an outlook on the full execution of the
extended algorithms (Algorithm \ref{alg:testRedundancy} and Algorithm
\ref{alg:complete}) over a set of filtering rules based on Table
\ref{tab:frules} -- whose main overlaps have been previously shown in
Figure \ref{fig:mitivationExample}.

\begin{center}
\begin{minipage}{7cm}
  $/*phase~1, step = 1*/$\\
  $R_1: (s \in [1,30] \land d \in [20,45]) \rightarrow deny$ \\
  $R_2: (s \in [31,60] \land d \in [25,35]) \rightarrow accept$ \\
  $R_3: (s \in [40,70] \land d \in [20,45]) \rightarrow accept$ \\
  $R_4: (s \in [15,45] \land d \in [25,30]) \rightarrow deny$ \\
  $R_5: (s \in [31,45] \land d \in [20,40]) \rightarrow accept$
\end{minipage}
\end{center}

\begin{center}
\begin{minipage}{7cm}
  $/*phase~1, step = 2, 3, 4 */$\\
  $R_1: (s \in [1,30] \land d \in [20,45]) \rightarrow deny$ \\
  $R_2: (s \in [31,60] \land d \in [25,35]) \rightarrow accept$ \\
  $R_3: (s \in [40,70] \land d \in [20,45]) \rightarrow accept$ \\
  $R_4: (s \in [15,30] \land d \in [25,30]) \rightarrow deny$ \\
  $R_5: (s \in [31,45] \land d \in [20,40]) \rightarrow accept$
\end{minipage}
\end{center}

\begin{center}
\begin{minipage}{7cm}
  $/*phase~2, step = 1 */$\\
  $/*testRedundancy(R_1) = false */$\\
  $R_1: (s \in [1,30] \land d \in [20,45]) \rightarrow deny$ \\
  $R_2: (s \in [31,60] \land d \in [25,35]) \rightarrow accept$ \\
  $R_3: (s \in [40,70] \land d \in [20,45]) \rightarrow accept$ \\
  $R_4: \emptyset \rightarrow accept$ \\
  $R_5: (s \in [31,45] \land d \in [20,40]) \rightarrow accept$
\end{minipage}
\end{center}

\begin{center}
\begin{minipage}{7cm}
  $/*phase~2, step = 2 */$\\
  $/*testRedundancy(R_2) = true */$\\
  $R_1: (s \in [1,30] \land d \in [20,45]) \rightarrow deny$ \\
  $R_2: \emptyset \rightarrow accept$ \\
  $R_3: (s \in [40,70] \land d \in [20,45]) \rightarrow accept$ \\
  $R_4: \emptyset \rightarrow accept$ \\
  $R_5: (s \in [31,45] \land d \in [20,40]) \rightarrow accept$
\end{minipage}
\end{center}

\begin{center}
\begin{minipage}{7cm}
  $/*phase~2, step = 3 */$\\
  $/*testRedundancy(R_3) = false */$\\
  $R_1: (s \in [1,30] \land d \in [20,45]) \rightarrow deny$ \\
  $R_2: \emptyset \rightarrow accept$ \\
  $R_3: (s \in [40,70] \land d \in [20,45]) \rightarrow accept$ \\
  $R_4: \emptyset \rightarrow accept$ \\
  $R_5: (s \in [31,39] \land d \in [20,40]) \rightarrow accept$
\end{minipage}
\end{center}

\begin{center}
\begin{minipage}{7cm}
  $/*phase~2, step = 4,5 */$\\
  $/*testRedundancy(R_4) = false */$\\
  $/*testRedundancy(R_5) = false */$\\
  $R_1: (s \in [1,30] \land d \in [20,45]) \rightarrow deny$ \\
  $R_2: \emptyset \rightarrow accept$ \\
  $R_3: (s \in [40,70] \land d \in [20,45]) \rightarrow accept$ \\
  $R_4: \emptyset \rightarrow accept$ \\
  $R_5: (s \in [31,39] \land d \in [20,40]) \rightarrow accept$
\end{minipage}
\end{center}

\begin{center}
\begin{minipage}{7cm}
  $/* resulting~rules */$\\
  $R_1: (s \in [1,30] \land d \in [20,45]) \rightarrow deny$ \\
  $R_3: (s \in [40,70] \land d \in [20,45]) \rightarrow accept$ \\
  $R_5: (s \in [31,39] \land d \in [20,40]) \rightarrow accept$
\end{minipage}
\end{center}

\begin{figure*}[hbtp]
 \begin{center}
    \subfigure[Best case example\label{fig:BestCase}]{
      \epsfig{file=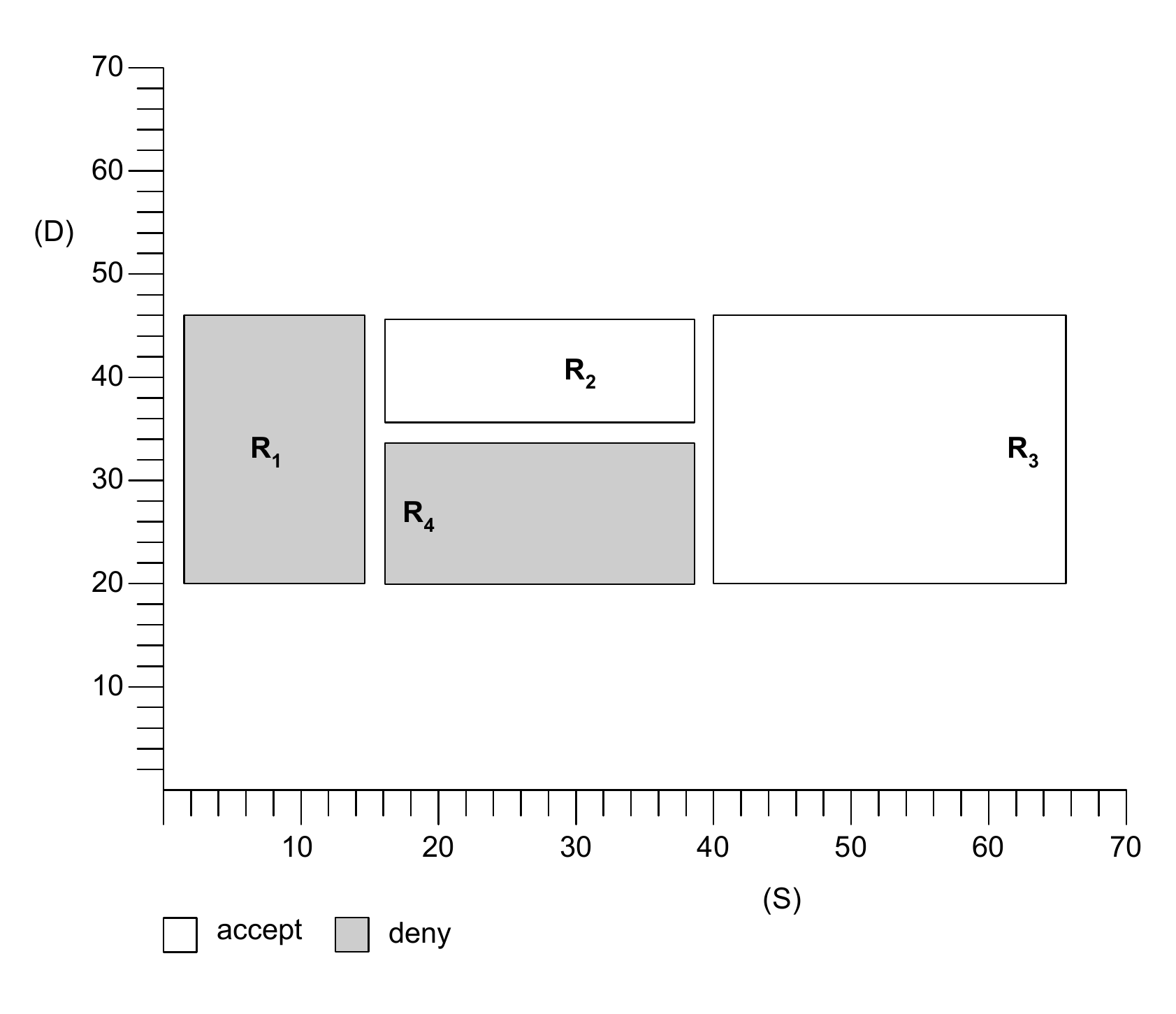, width=5cm}
    }
    \subfigure[Normal case example\label{fig:NormalCase}]{
      \epsfig{file=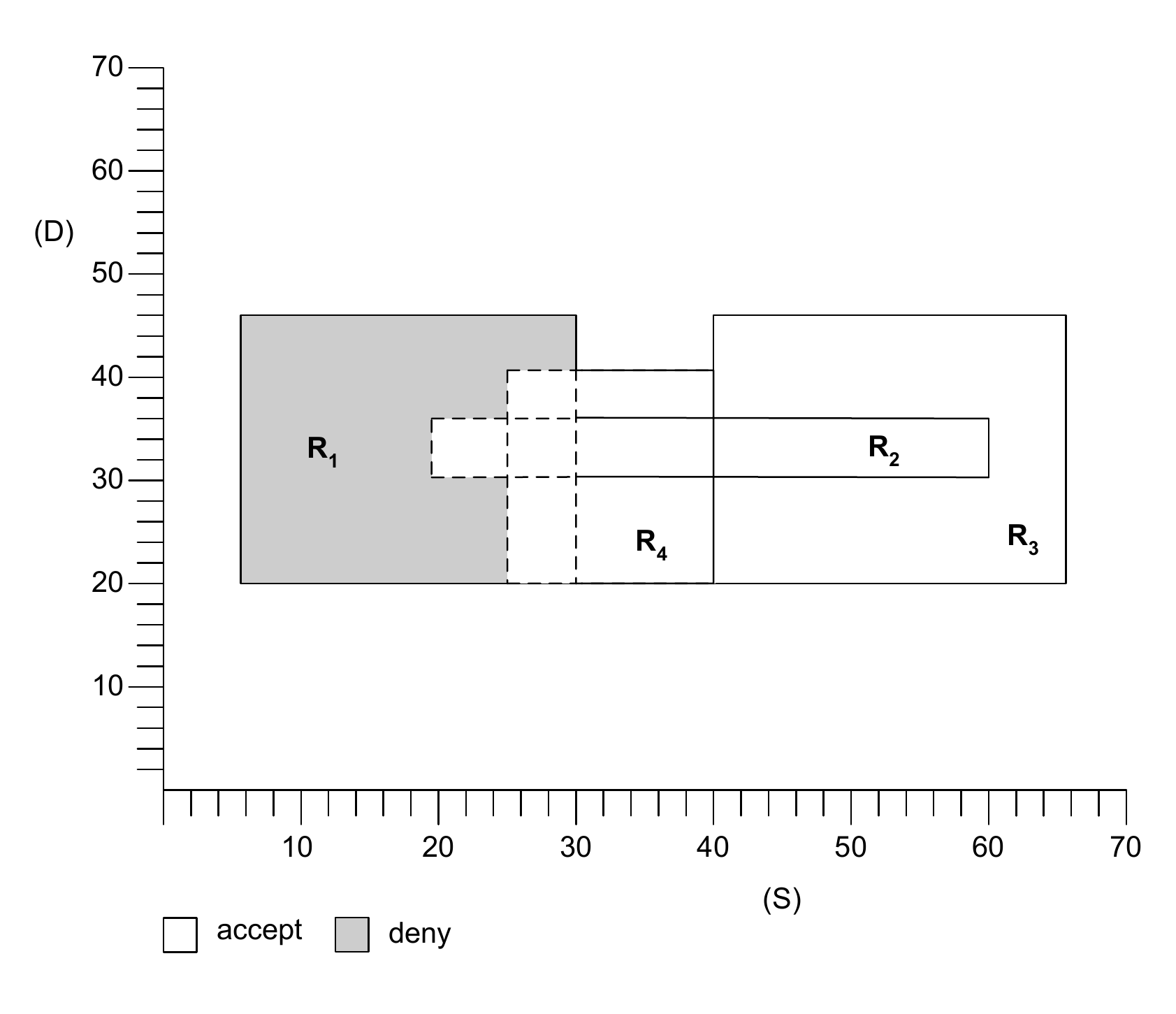, width=5cm}
    }
    \subfigure[Worst case example\label{fig:WorstCase}]{
      \epsfig{file=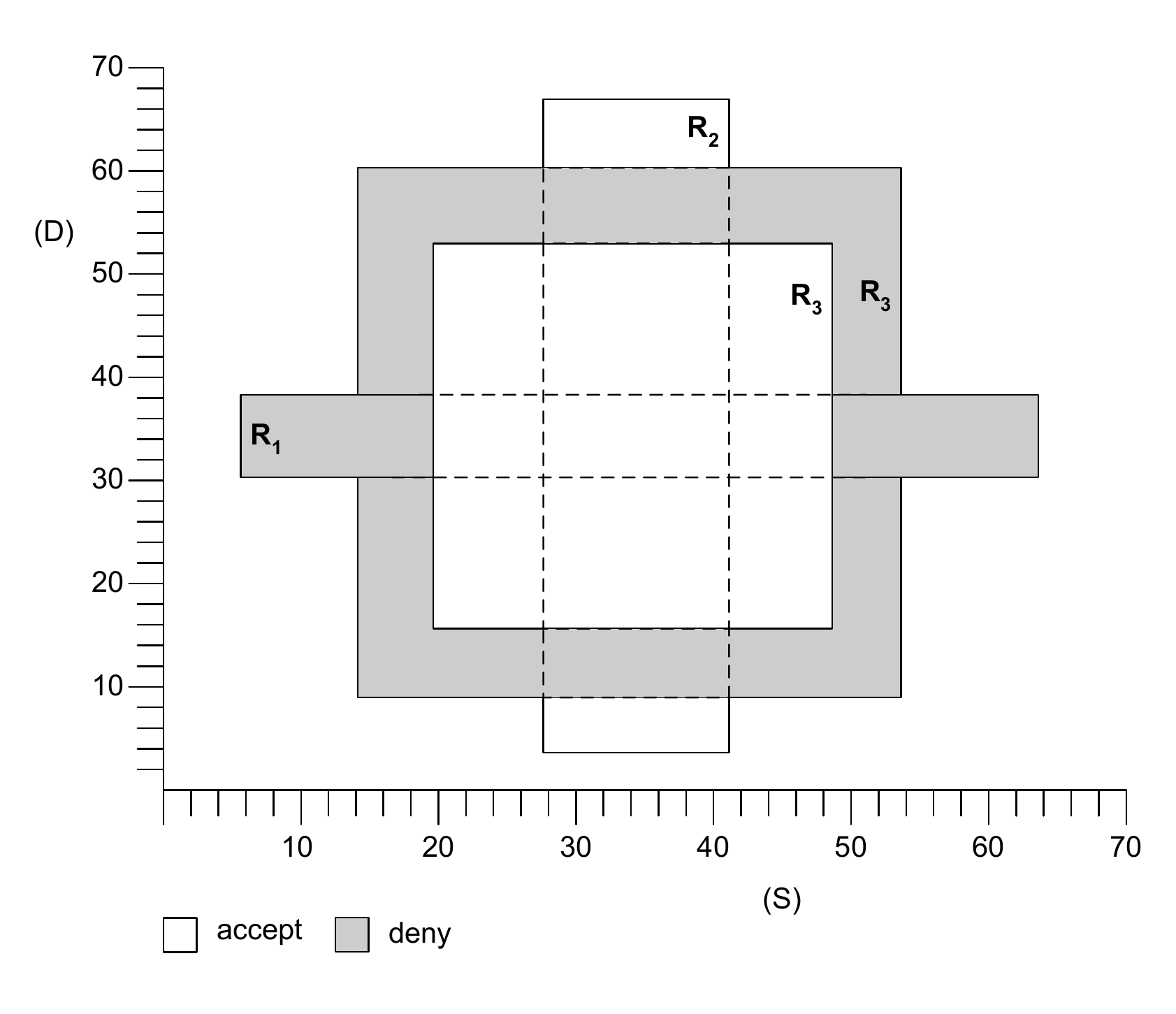, width=5cm}
    }
    \caption{Best, normal and worst ruleset examples}
    \label{fig:examples}
 \end{center}
\end{figure*}

To conclude, let us recall that the following two warnings will
notice the security officer to the discovering of both shadowing and
redundancy errors, in order to verify the correctness of the whole
detection and transformation process:

\begin{center}
\begin{minipage}{7cm}
  $/*warnings*/$\\
  $R_2[redundancy]=true$\\
  $R_4[shadowing]=true$\\
\end{minipage}
\end{center}

\subsection{Correctness of the Algorithms}

\begin{thm}\label{thm:four}
  Let $R$ be a set of filtering rules and let $Tr'(R)$ be the
  resulting filtering rules obtained by applying Algorithm
  \ref{alg:complete} to $R$. Then $R$ and $Tr'(R)$ are
  equivalent.\footnote{A set of proofs to validate the theorems of
    this section is provided in \cite{phoenix}.}
\end{thm}

\begin{thm}\label{thm:five}
  Let $R$ be a set of filtering rules and let $Tr'(R)$ be the
  resulting filtering rules obtained by applying Algorithm
  \ref{alg:complete} to $R$. Then ordering the rules in $Tr'(R)$ is no
  longer relevant.
\end{thm}

\begin{thm}\label{thm:six}
  Let $R$ be a set of filtering rules and let $Tr'(R)$ be the
  resulting filtering rules obtained by applying Algorithm
  \ref{alg:complete} to $R$. Then $Tr'(R)$ is free from both shadowing
  and redundancy.
\end{thm}

\subsection{Complexity of the Algorithms}

In the worst case, Algorithm 4 presented in this paper may generate a
large number of rules. If we have 2 rules with $p$ attributes, the
second rule can be replaced by $p$ new rules in the worst case,
leading to $p + 1$ rules.

If we now assume that we have $n$ rules ($n > 2$) with $p$ attributes,
then each rule except the first one can be replaced by $p$ new rules
in the first rewriting step of the algorithm. In the second rewriting
step, the $p$ rules that replace the second rule are combined with the
$p$ rules that replace rules 3 to $n$. Thus, each rule from 3 to $n$
can be replaced by $p^2$ new rules. In the third step, the $p^2$ rules
corresponding to rule 3 are combined with the $p^2$ rules
corresponding to rules 4 to $n$. We can show that this may lead to
$p^3$ new rules. And so on.

So, in the worst case, if we have $n$ rules ($n > 2$) with $p$
attributes, then we can obtain $1 + p + p^2 + \ldots + p^{n-1}$ rules
when applying Algorithm 4, that is $\frac{p^n - 1}{p - 1}$ rules.

Thus, complexity of Algorithm \ref{alg:exclusion} is very high.
However, in all the experimentations we have done (cf.~ Section
\ref{sec:evaluation}), we were always very far from the worst case.
First, because only attributes source and destination may
significantly overlap and exercice a bad influence on the algorithm
complexity. Other attributes, protocoles and source and destination
port numbers, are generally equal or completely different when
combining configuration rules. Second, administrators generally use
overlapping rules in their firewall configurations to represent rules
that may have {\em exceptions}. This situation is closer to the normal
case presented in Figure~\ref{fig:examples} than to the worst case.
Third, when shadowing or redundancy situations are discovered by the
algorithm, some rules are removed -- which significantly reduces the
algorithm complexity.

\section{Performance Evaluation}
\label{sec:evaluation}

We have implemented the algorithms described in Section
\ref{sec:proposal} in a software prototype called 
\href{https://github.com/jgalfaro/mirrored-mirage}{MIRAGE
(MIsconfiguRAtion manaGEr)}. MIRAGE has been developed using PHP, a
general-purpose scripting language that is especially suited for web
services development and can be embedded into HTML for the
construction of client-side GUI based applications \cite{php}. MIRAGE
can be locally or remotely executed by using a HTTP server (e.g.,
Apache server over UNIX or Windows setups) and a web browser.

In this section, we present an evaluation of the performance of MIRAGE
applying the set of detection and removal algorithms over the
filtering rules of a simulated IPv4 network.

Inspired by the experiments done in \cite{al-shaer04,al-shaer05}, we
simulated in a first phase several sets of IPv4 filtering policies,
according to the three following security officer profiles: beginner,
intermediate, and expert -- where the probability to have overlaps
between rules increases from 5\% to 90\%. Then, we processed in a
second phase all these sets of filtering rules within our prototype,
in order to evaluate its performance and scalability.

\begin{figure}[htbp]
 \centering
 \epsfig{file=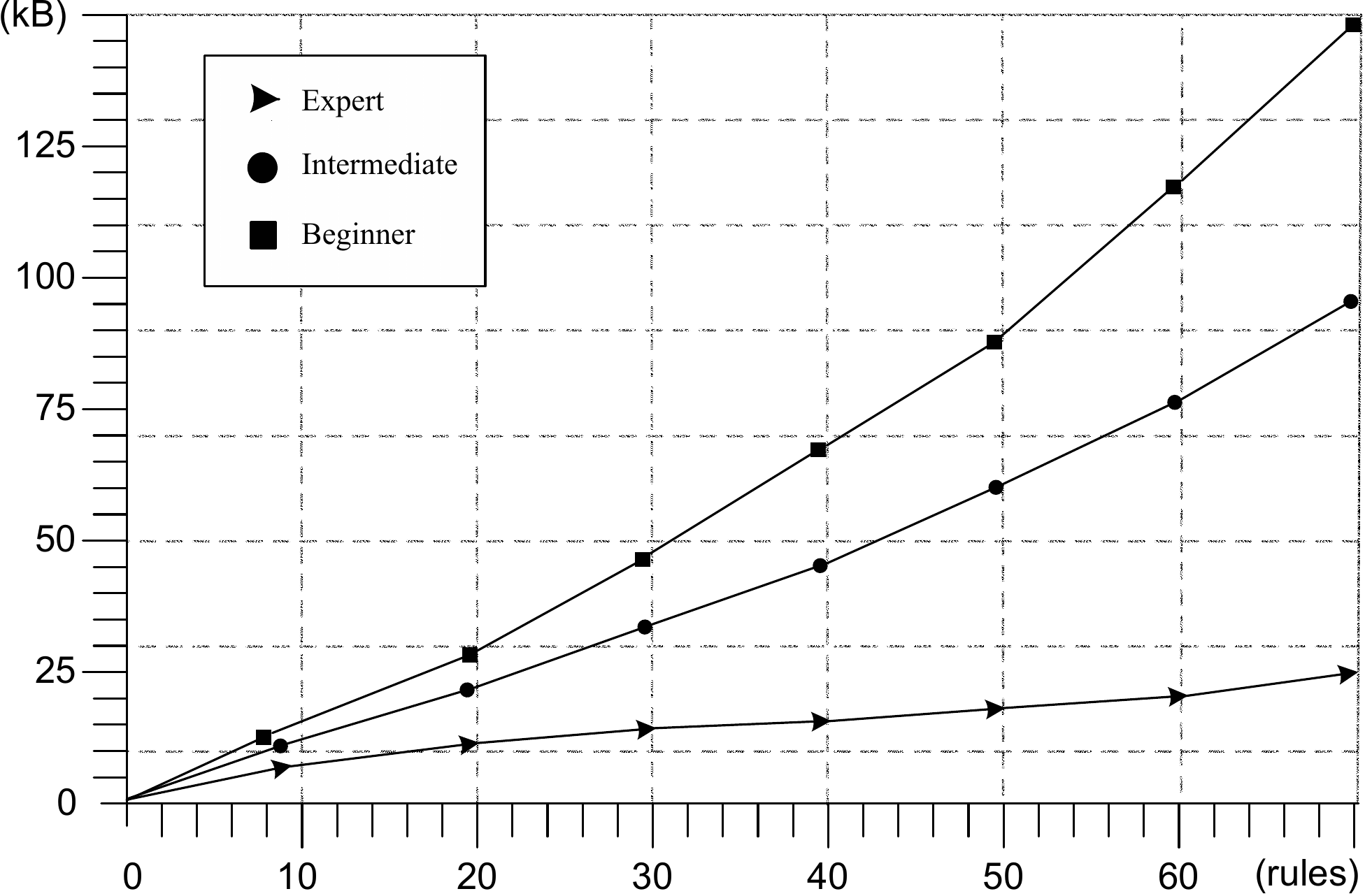, width=7.5cm}
 \caption{Memory space evaluation}
 \label{fig:memoryEvaluation}
\end{figure}

\begin{figure}[htbp]
 \centering
 \epsfig{file=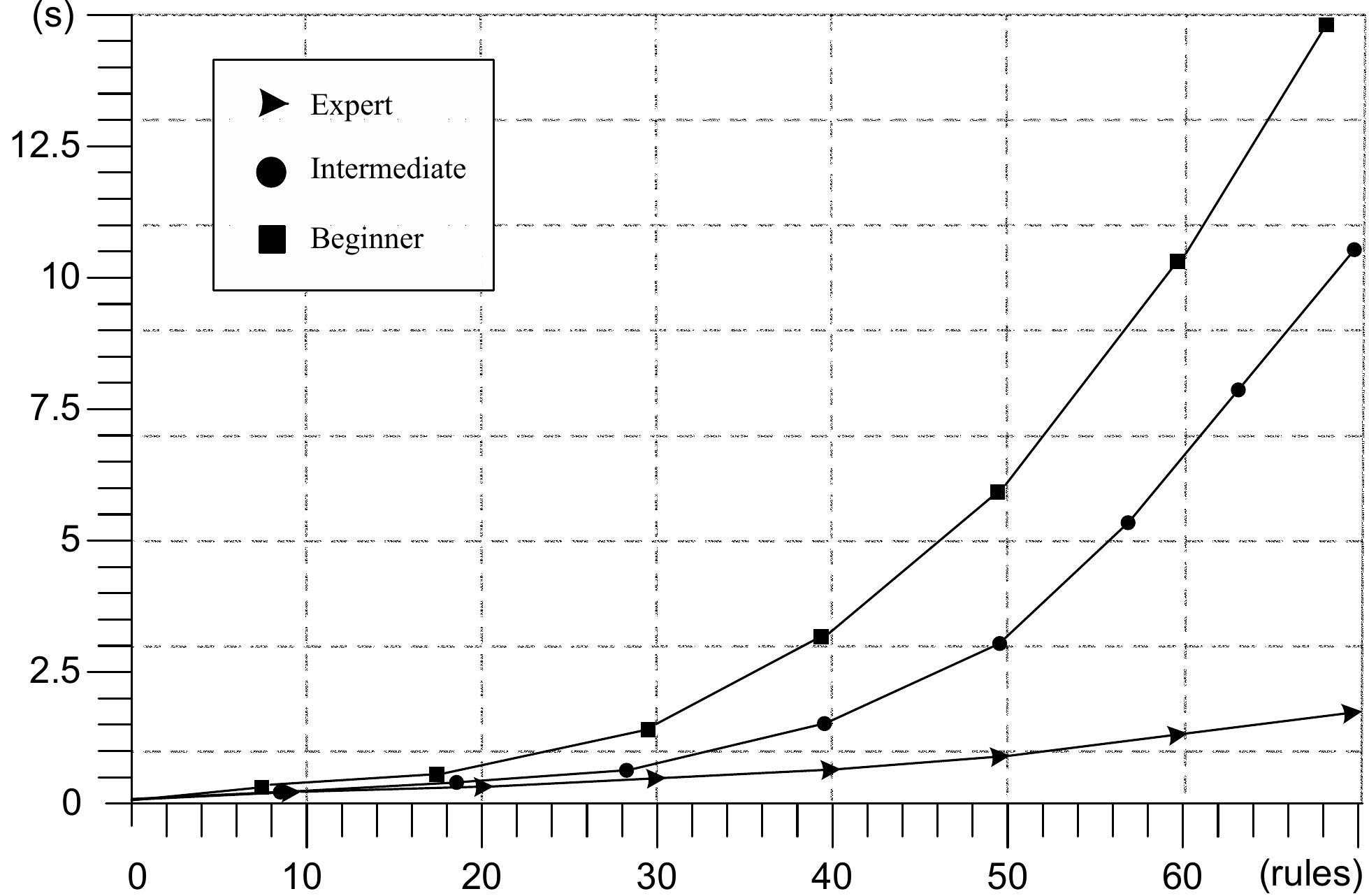, width=7.5cm}
 \caption{Processing time evaluation}
 \label{fig:processingEvaluation}
\end{figure}

The whole of these experiments were carried out on an Intel-Pentium M
1.4 GHz processor with 512 MB RAM, running Debian GNU/Linux 2.6.8, and
using Apache/1.3 with PHP/4.3 interpreter configured. During these
experiments, we measured the memory space and the processing time
needed to perform algorithms \ref{alg:exclusion},
\ref{alg:testRedundancy}, and \ref{alg:complete}. The results of these
measurements are plotted in Figure \ref{fig:memoryEvaluation} and
Figure \ref{fig:processingEvaluation}. Although the plots reflect
strong memory and process time requirements, we consider they are
reasonable for off-line analysis, since it is
not part of the critical performance of a firewall.\\

\section{Conclusions}
\label{sec:conclusions}

There are two ways to set a firewall configuration free of errors. A
first approach is to apply a formal security model -- such as the
formal model presented in \cite{fast} -- to express the security
policy of the access control for the network, and to generate the
specific syntax for each given firewall from this formal policy -- for
instance, by using XSL transformations from the formal policy to
generate specific Netfilter configuration rules \cite{netfilter}. The
main advantage of this approach is the great confidence we have in the
conformity of the formal policy, and its translation into a specific
firewall configuration. Nevertheless, although a great number of
errors is avoided when using this formal approach, it is still not
ensured that all the possible errors are discarded.

A second approach -- as the one presented in this paper -- is to apply
an audit process to the set of filtering rules of a given firewall --
which expresses a specific network security policy -- in order to
detect configuration errors and to properly eliminate them. In our
case, the audit process is based on the existence of relationships
between the condition attributes of the filtering rules, such as
coincidence, disjunction, and inclusion. Then, our proposal uses a
transformation process which derives from an initial set of rules --
potentially misconfigured -- to an equivalent one which is completely
free of misconfiguration.

Some other advantages of our approach are the following. First of all,
our transformation process verify that the resulting rules are
completely independent between them. Otherwise, each redundant or
shadowed rule considered as useless during the process is removed from
the configuration. On the other hand, the discovering process provides
an evidence of error to the administration console. This way, the
security officer can check whether the security policy is consistent,
in order to verify the correctness of the process.

The complete independence between rules, moreover, enables the
possibility to perform a second rewriting of rules in a positive
manner -- only permissions -- or in a negative manner -- only
prohibitions. After performing this second transformation, the
security officer will have a clear view of the accepted traffic --
when positive rewriting -- or the rejected traffic -- when negative
rewriting.

Regarding a possible increase of the initial number of filtering
rules, due to the applying of Algorithm \ref{alg:exclusion}, it is
only significant whether the associated parsing algorithm of the
firewall depends on the number of rules. In this case, an increase in
such a parameter may degrade the performance of the firewall.
Nonetheless, this is not a disadvantage since the use of a parsing
algorithm independent of the number of rules becomes the best solution
as much for our proposal as for the current deployment of firewall
technologies. The set pruning tree algorithm is a proper example,
because it only depends on the number and size of attributes to be
parsed, not the number of rules \cite{paul00}.

The implementation of the algorithms in a software prototype
demonstrate the practicability of our work. We shortly discussed this
implementation, based on a general-purpose scripting language
\cite{php}, and presented an evaluation of its performance. Although
the experimental results show that our algorithms have strong memory
and process time requirements, we believe that these requirements are
reasonable for off-line analysis, since it is not part of the critical
performance of a firewall.

As future work we are considering to extend our proposal to a more
complex firewall setup. The work stated in this paper is based on the
hypothesis that only one firewall ensures the network access control.
More investigation has to be done when this role is assigned to more
than one network security component, that is a distributed access
control. Indeed, in particular, redundancy will not systematically be
considered as an error \cite{al-shaer04}. It may be suited in order to
avoid inconsistent decisions between firewalls used in the same
security architecture to control the access to different zones.

In parallel to this work, we also study the anomaly problems of
security rules in the case where the security architecture includes
firewalls as well as IDS (Intrusion Detection Systems). The objective
is to avoid redundant or shadowed filtering or/and alerting rules.
Indeed, there is a real similarity between the parameters of a
filtering rule and those of an alerting rules (signatures) so that we
can apply algorithms presented in both Section
\ref{sec:detectingShadowing} and Section \ref{sec:extendedDetection}.
Of course, this will depend on whether the firewall is the first
security component in the security architecture that the packets
encounter or it acts after the detection intrusion component.

\section*{Acknowledgements}

This work was supported by funding from the French ministry of
research, under the \textit{ACI DESIRS} project, the Spanish
Government project \textit{TIC2003-02041}, and the Catalan Government
grants \textit{2003FI126} and \textit{2005BE77}.


\newpage

\appendix

\section{Correctness Proofs\label{ap:proofs}}

\begin{proof}\textbf{of Lemma \ref{lem:equivalence}}~~~
  Let us assume that:

  \begin{center}
    \begin{minipage}{6cm}
      $R_i[condition] = A_1 \wedge A_2 \wedge ... \wedge A_p$, and \\
      $R_j[condition] = B_1 \wedge B_2 \wedge ... \wedge B_p$.
    \end{minipage}
  \end{center}

  \noindent
  If $(A_1 \cap B_1) = \emptyset$ or $(A_2 \cap B_2) = \emptyset$ or
  \ldots or $(A_p \cap B_p) = \emptyset$ then $exclusion(R_j, R_i)
  \leftarrow R_j$. Hence, to prove the equivalence between
  $\{R_i,R_j\}$ and $\{R_i,R'_j\}$ is trivial in this case.\\

  \noindent
  Let us now assume that:
  \begin{center}
    \begin{minipage}{6cm}
      $(A_1 \cap B_1) \neq \emptyset$ and
      $(A_2 \cap B_2) \neq \emptyset$ and
       ... and  $(A_p \cap B_p) \neq \emptyset$.
    \end{minipage}
  \end{center}

  \noindent
  If we apply filtering rules $\{R_i,R_j\}$ where $R_i$ comes before
  $R_j$, then rule $R_j$ applies to a given packet if this packet
  satisfies $R_j[condition]$ but not $R_i[condition]$ (since rule
  $R_i$ applies first). Therefore, notice that $R_j[condition] -
  R_i[condition]$ is equivalent to:\\

  \begin{center}
    \begin{minipage}{6.9cm}
        $(B_1 - A_1) \wedge B_2 \wedge...\wedge B_p$ or\\
        $(A_1 \cap B_1) \wedge (B_2 - A_2) \wedge...\wedge B_p$ or\\
        $(A_1 \cap B_1) \wedge (A_2 \cap B_2) \wedge (B_3 - A_3) \wedge...\wedge B_p$ or\\
        $ ...$ \\
        $(A_1 \cap B_1) \wedge ... \wedge (A_{p-1} \cap B_{p-1}) \wedge (B_p -
        A_p)$
    \end{minipage}
    \vspace{-0.25cm}
  \end{center}
  ~~\\

  \noindent
  which corresponds to condition of rule $R'_j = exclusion(R_j, R_i)$.
  This way, if rule $R_j$ applies to a given packet in $\{R_i,R_j\}$,
  then rule $R'_j$ also applies to this packet in $\{R_i,R'_j\}$.\\

  \noindent
  Conversely, if rule $R'_j$ applies to a given packet in
  $\{R_i,R'_j\}$, then this means this packet satisfies
  $R_j[condition]$ but not $R_i[condition]$. So, it is clear that rule
  $R_j$ also applies to this packet in $\{R_i,R_j\}$.\\

  \noindent
  Since in Algorithm \ref{alg:exclusion} $R'_j[decision]$ becomes
  $R_j[decision]$, this enables to conclude that $\{R_i,R_j\}$ is
  equivalent to $\{R_i,R'_j\}$.\qed
\end{proof}

\begin{proof}\textbf{of Theorem \ref{thm:equivalence}}~~~
  Notice that if $R$ is a set of filtering rules,
  then $Tr(R)$ is obtained by recursively applying transformation
  $exclusion(R_j, R_i)$ when rule $R_i$ comes before rule $R_j$, which
  preserves the equivalence at each step of the transformation,
  previously proved for Lemma \ref{lem:equivalence}.\qed
\end{proof}

\begin{proof}\textbf{of Lemma \ref{lem:simultaneousness}}~~~
  Notice that rule $R'_j$ only applies when rule $R_i$ does not apply.
  Thus, if rule $R'_j$ comes before rule $R_i$, this will not change
  the final decision since rule $R'_j$ only applies to packets that do
  not match rule $R_i$. \qed
\end{proof}

\begin{proof}\textbf{of Theorem \ref{thm:ordering}}~~~
  For any pair of rules $R_i$ and $R_j$ such that $R_i$ comes before
  $R_j$, $R_j$ is replaced by a rule $R'_j$ obtained by recursively
  replacing $R_j$ by $exclusion(R_j,R_k)$ for any $k < j$.\\

  \noindent
  Then, by recursively applying Lemma \ref{lem:simultaneousness}, it
  is possible to commute rules $R'_i$ and $R'_j$ in $Tr(R)$ without
  changing the final decision. \qed
\end{proof}

\begin{proof}\textbf{of Theorem \ref{thm:freedom}}~~~
  Notice that, in $Tr(R)$, each rule is independent of all other
  rules. Thus, if we consider a rule $R_i$ in $Tr(R)$ such that
  $R_i[condition] \neq \emptyset$, then this rule will apply to any
  packet that satisfies $R_i[condition]$. Hence, this rule is not
  shadowed.\\

  \noindent
  Similarly, rule $R_i$ is not redundant because if we remove this
  rule, since this rule is the only one that applies to packets that
  satisfy $R_i[condition]$, then the filtering decision will change if
  we remove rule $R_i$ from $Tr(R)$. \qed
\end{proof}

\begin{proof}\textbf{of Theorem \ref{thm:four}}~~~
  Let $Tr'_1(R)$ be the set of rules obtained after applying the first
  phase of Algorithm \ref{alg:complete}. Since $Tr'_1(R)$ is derived
  from $R$ by applying $exclusion(R_j,R_i)$ (cf.~Algorithm
  \ref{alg:exclusion}) to some rules $R_j$ in $R$, it is
  straightforward, from Lemma \ref{lem:equivalence}, to conclude that
  $Tr'_1(R)$ is equivalent to $R$.\\

  \noindent
  Hence, let us now move to the second phase of Algorithm
  \ref{alg:complete}. Let us consider a rule $R_i$ such that
  $testRedundancy(R_i)$ (cf.~Algorithm \ref{alg:testRedundancy}) is
  $true$. This means that $R_i[condition]$ can be derived by
  conditions of a set of rules $S$ with the same decision and that
  come after in order than rule $R_i$.\\

  \noindent
  Since every rule $R_j$ with a decision different from the one of
  rules in $S$ has already been excluded from rules of $S$ in the
  first phase of the Algorithm, we can conclude that rule $R_i$ is
  definitely redundant and can be removed without changing the final
  filtering decision. This way, we conclude that Algorithm
  \ref{alg:complete} preserves equivalence in this case.\\

  \noindent
  On the other hand, if $testRedundancy(R_i)$ is $false$, then
  transformation consists in applying $exclusion(R_j,R_i)$ to some
  rules $R_j$ which also preserves equivalence. Thus, in both cases,
  $Tr'(R)$ is equivalent to $Tr'_1(R)$ which, in turn, is equivalent
  to $R$. \qed
\end{proof}

\begin{proof}\textbf{of Theorem \ref{thm:five} and Theorem \ref{thm:six}}~~~
  As stated out in the proof of both Theorem \ref{thm:ordering} and
  Theorem \ref{thm:freedom}, once shadowed and redundant rules have
  been removed, every rule $R_i$ is replaced by $exclusion(R_i,R_j)$
  where $j < i$.\\

  \noindent
  Therefore, a similar reasoning enables to prove both Theorem
  \ref{thm:five} and Theorem \ref{thm:six}. \qed
\end{proof}

\end{document}